\documentstyle[11pt,epsfig,psfrag,wrapfig]{article}
   \newcommand{\la}{\langle}
   \newcommand{\ra}{\rangle}
   \newcommand{\be}{\begin{equation}}
   \newcommand{\ee}{\end{equation}}
   \newcommand{\ba}{\begin{eqnarray}}
   \newcommand{\ea}{\end{eqnarray}}
   \newcommand{\di}{{\rm d}}
\begin{document}
%
\title{Azimuthal asymmetries and Collins analyzing power \thanks{Presented 
by A.~V.~Efremov at DIS2002, Krak\'ow, Poland, 30 April to 4 May 2002.}}
\author{A.~V.~Efremov$^a$, K.~Goeke$^b$ and P.~Schweitzer$^c$\\
\footnotesize
$^a$ Joint Institute for Nuclear Research, Dubna, 141980 Russia\\
\footnotesize
$^b$ Institute for Theoretical Physics II, Ruhr University Bochum, Germany\\
\footnotesize
$^c$ Dipartimento di Fisica Nucleare e Teorica, Universit\`a di Pavia, Italy}
\date{}
\maketitle
	\vspace{-9cm} \begin{flushright} RUB-TP2-10-02
	\end{flushright}\vspace{7cm}
\begin{abstract}
Spin azimuthal asymmetries in pion electro-production in deep inelastic
scattering off longitudinally polarized protons, measured by HERMES,
are well reproduced theoretically with no adjustable parameters.
Predictions for azimuthal asymmetries for a longitudinally polarized deuteron
target are given. 
The z-dependence of the Collins fragmentation function is extracted.
The first information on $e(x)$ is extracted from CLAS $A_{LU}$ asymmetry.
\end{abstract}
%
\section{Introduction}
Recently azimuthal asymmetries have been observed in pion electro-production
in semi inclusive deep-inelastic scattering off longitudinally (with respect
to the beam) \cite{hermes,hermes-pi0} and transversely polarized protons
\cite{bravardis99}.
These asymmetries contain information on the T-odd ``Collins'' fragmentation
function $H_1^{\perp a}(z)$ and on the transversity distribution $h_1^a(x)$
\cite{transversity}\footnote{We use the notation of
    Ref.\cite{muldt,Mulders:1996dh} with $H_1^{\perp}(z)$
    normalized to $\la P_{h\perp}\ra$ instead of $M_h$.}.
$H_1^{\perp a}(z)$ describes the left-right asymmetry in fragmentation of
transversely polarized quarks into a hadron \cite{muldt,Mulders:1996dh,collins}
(the "Collins asymmetry"),  and $h_1^a(x)$ describes the distribution of
transversely polarized quarks in nucleon \cite{transversity}.
Both $H_1^{\perp a}(z)$ and $h_1^a(x)$ are twist-2, chirally odd, and not
known experimentally.  Only recently experimental indications to
$H_1^{\perp}$ in $e^+e^-$-annihilation have appeared \cite{todd},
while the HERMES and SMC data \cite{hermes,hermes-pi0,bravardis99}
provide first experimental indications to $h_1^a(x)$.

Here we explain the observed azimuthal asymmetries \cite{hermes,hermes-pi0}
and predict pion and kaon asymmetries from a deuteron target for HERMES by
using information on $H_1^{\perp}$ from DELPHI \cite{todd} and
the predictions for the transversity distribution $h_1^a(x)$ from
the chiral quark-soliton model ($\chi$QSM) \cite{h1-model}.
Our analysis is free of any adjustable parameters.
Moreover, we use the model prediction for $h_1^a(x)$ to extract
$H_1^\perp(z)$ from the $z$-dependence of HERMES data.
For more details and complete references see
Ref.\cite{Efremov:2000za,Efremov:2001cz,Efremov:2001ia}.
Finally, using the new information on $H_1^\perp(z)$, we extract
the twist-3 distribution $e^a(x)$ from very recent CLAS data \cite{ALU}.

\section{Transversity distribution and Collins fragmentation function}
\label{sect-h1-H1perp}
The $\chi$QSM is a quantum field-theoretical relativistic model with
explicit quark and antiquark degrees of freedom. This allows an
unambiguous identification of quark {\sl and} antiquark distributions
in the nucleon, which satisfy all general QCD requirements
due to the field-theoretical nature of the model \cite{DPPPW96}.
The results of the parameter-free calculations for unpolarized and helicity
distributions agree within $(10 - 20)\%$ with parameterizations,
suggesting a similar reliability of the model prediction for
$h_1^a(x)$ \cite{h1-model}.

$H_1^{\perp}$ is responsible in $e^+e^-$ annihilation for a specific
azimuthal asymmetry of a hadron in a jet around the axis in direction
of the second hadron in the opposite jet \cite{muldt}.  This asymmetry
was probed using the DELPHI data collection \cite{todd}.  For the
leading particles in each jet of two-jet events, averaged over quark flavors,
the most reliable value of the analyzing power is given by $(6.3\pm 2.0)\%$. 
However, the larger ``optimistic'' value is not excluded
\be\label{apower}
    	\left|{\la H_1^{\perp}\ra\over\la D_1\ra}\right| =(12.5\pm 1.4)\% 
	\ee
with unestimated but presumably large systematic errors.

\section{The azimuthal asymmetry}
In \cite{hermes,hermes-pi0} the cross section for $l\vec{p}\rightarrow l'\pi X$
was measured in dependence of the azimuthal angle $\phi$, i.e the angle
between lepton scattering plane and the plane defined by momentum of virtual
photon ${\bf q}$ and momentum ${\bf P}_{\!h}$ of produced pion.
The twist-2 and twist-3 azimuthal asymmetries read
\cite{Mulders:1996dh}\footnote{Note a sign-misprint in Eq.(115) of
    \cite{Mulders:1996dh} for the $\sin\phi$-term Eq.(\ref{AUL1}).
    It was corrected in Eq.(2) of \cite{Boglione:2000jk}.
    The conventions in Eqs.(\ref{AUL2}--\ref{AUT}) agree with
    \cite{hermes,hermes-pi0}: Target polarization opposite to beam
    is positive, and $z$ axis is parallel to ${\bf q}$
    (in \cite{Mulders:1996dh} it is anti-parallel). }
\ba
\label{AUL2}
    A_{UL}^{\sin2\phi}(x) &\propto&
    {\sum_a e_a^2h^{\perp(1)a}_{1L}(x)\la H^{\perp a/\pi}_1\ra}\biggl/
    {\sum_a e_a^2 f^a_1(x)\la D_1^{a/\pi}\ra}\;,  \\
\label{AUL1}
    A_{UL(1)}^{\sin\phi}(x) &\propto&
    \frac{M}{Q}
    {\sum_a e_a^2xh^a_L(x)\la H^{\perp a/\pi}_1\ra}\biggl/
    {\sum_a e_a^2 f^a_1(x)\la D_1^{a/\pi}\ra}\; , \\
\label{AUT}
    A_{UL(2)}^{\sin\phi}(x) &\propto&
    -\sin\theta_\gamma\cdot
    {\sum_a e_a^2 h^a_1(x)\la H^{\perp a/\pi}_1\ra}\biggl/
    {\sum_a e_a^2 f^a_1(x)\la D_1^{a/\pi}\ra}\;, \ea
with $\sin\theta_\gamma\approx 2x\sqrt{1-y}(M/Q)$ and
$A^{\sin\phi}_{UL}=A^{\sin\phi}_{UL(1)}+A^{\sin\phi}_{UL(2)}$.
In Eqs.(\ref{AUL2}-\ref{AUT}) the pure twist-3 terms are neglected.
The results of Ref.\cite{Dressler:2000hc} justify to use this
WW-type approximation in which
$xh_L = -{2}h_{1L}^{\perp(1)}=2x^2\int_x^1\di\xi\,h_1(\xi)/\xi^2$.

We assume isospin symmetry and favoured fragmentation for $D_1^a$ and
$H_1^{\perp a}$, i.e.
$D_1^\pi\equiv D_1^{u/\pi^+}\!\!=D_1^{d/\pi^-}\!\!=2D_1^{\bar u/\pi^0}$
etc. and $D_1^{\bar{u}/\pi^+}\!\!=D_1^{u/\pi^-}\!\!\simeq 0$ etc.
\section{Explaining, exploiting and predicting HERMES asymmetries}
\label{AUL-HERMES}
When using Eq.(\ref{apower}) to explain HERMES data,
we assume a weak scale dependence of the analyzing power.
We take $h_1^a(x)$ from the $\chi$QSM \cite{h1-model} and $f_1^a(x)$ from
Ref.\cite{GRV}, both LO-evolved to the average scale
$Q_{\rm av}^2=4\,{\rm GeV}^2$.

In Fig.1 HERMES data for $A_{UL}^{\sin\phi}(x)$, $A_{UL}^{\sin2\phi}(x)$
\cite{hermes,hermes-pi0} are compared with the results of our analysis.
We conclude that the azimuthal asymmetries obtained with $h_1^a(x)$ from the
$\chi$QSM \cite{h1-model} combined with the ``optimistic'' DELPHI result
Eq.(\ref{apower}) for the analyzing power are consistent with data.

\begin{figure}[h!]
    \mbox{\epsfig{figure=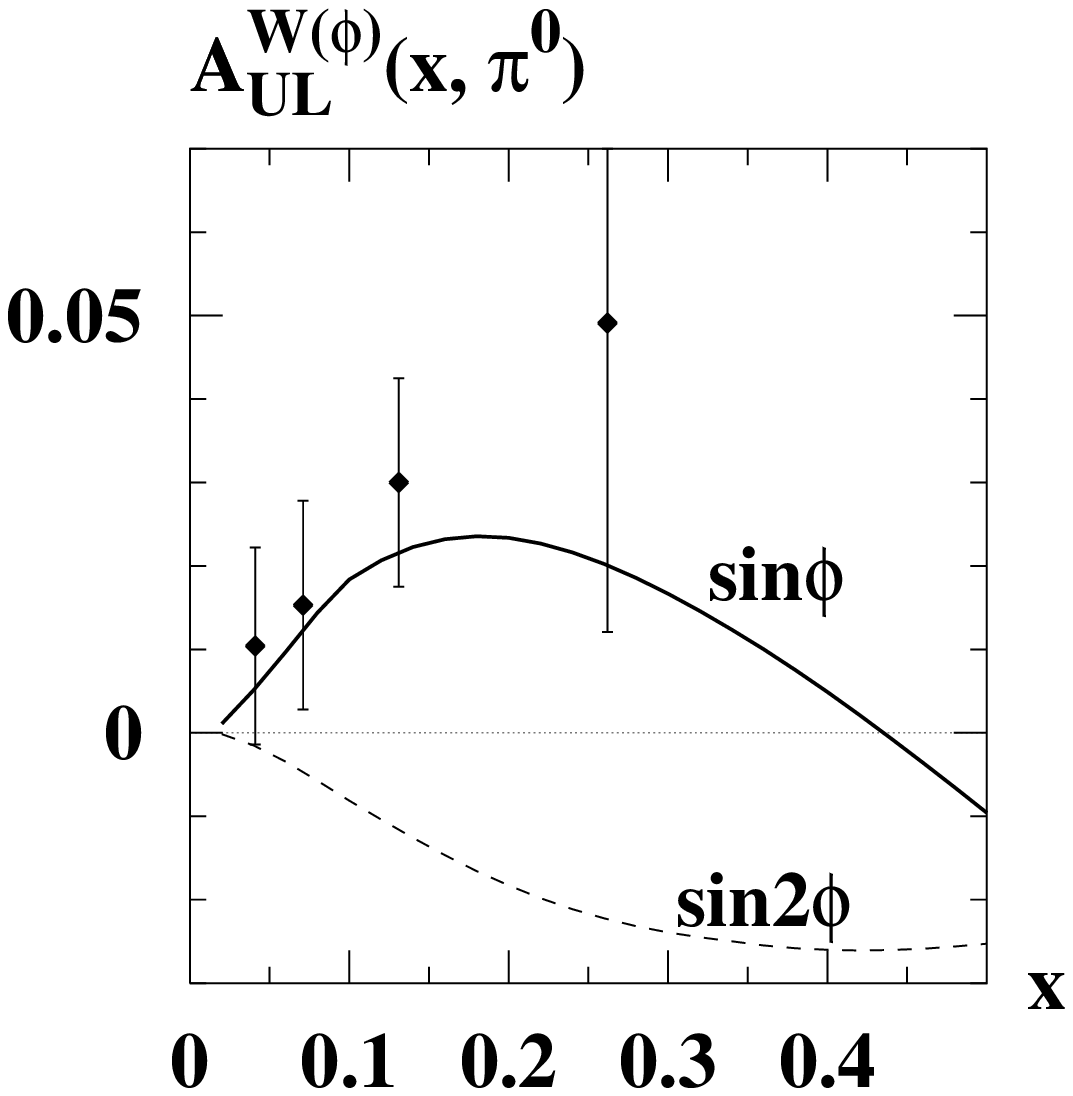,width=4cm,height=3.5cm}}
    \mbox{\epsfig{figure=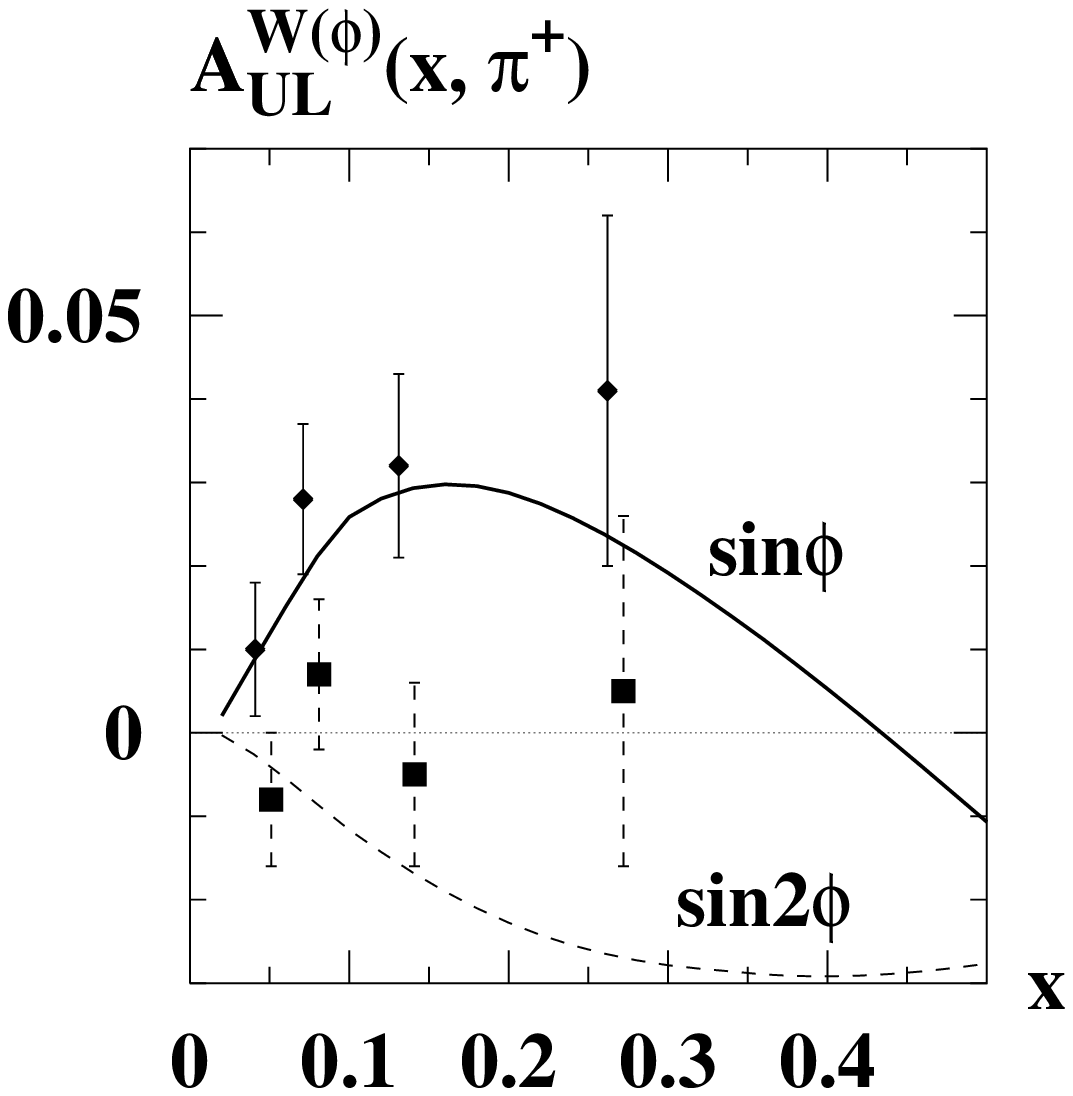,width=4cm,height=3.5cm}}
    \mbox{\epsfig{figure=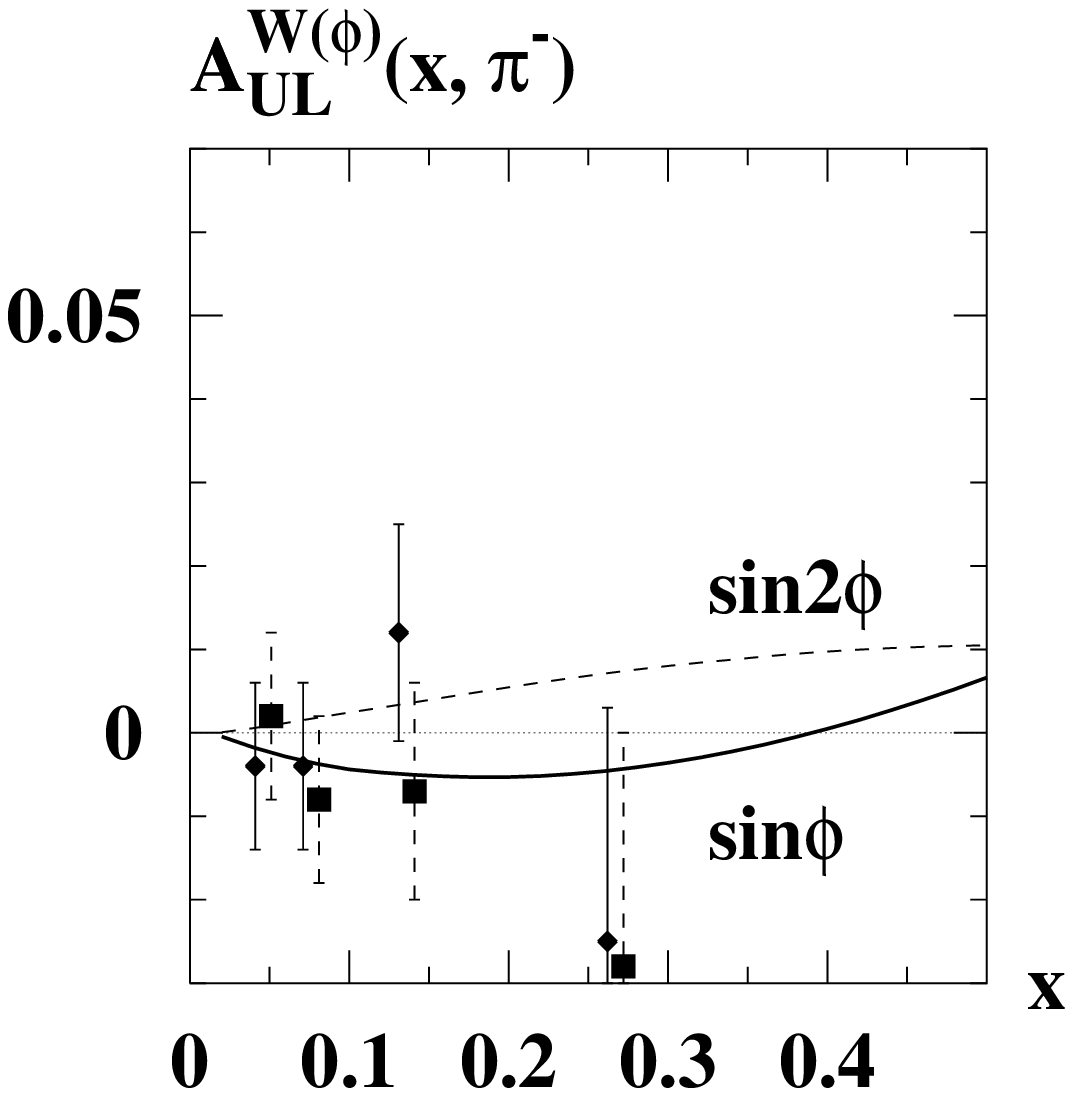,width=4cm,height=3.5cm}}
    \caption{Azimuthal asymmetries $A_{UL}^{W(\phi)}$ weighted by
    $W(\phi)=\sin\phi$, $\sin 2\phi$ for pions as function of $x$.
    Rhombus (squares) denote data for $A_{UL}^{\sin\phi}$
    ($A_{UL}^{\sin2\phi}$).}
\end{figure}

We exploit the $z$-dependence of HERMES data for $\pi^0$, $\pi^+$
azi\-muthal asymmetries to extract $H_1^\perp(z)/D_1(z)$.
For that we use the $\chi$QSM prediction for $h_1^a(x)$, which introduces a
model dependence of order $(10 - 20)\%$. The result is shown in Fig.2.
The data can be described by a linear fit
$H_1^\perp(z)=(0.33 \pm  0.06)zD_1(z)$. The average
${\la H_1^\perp\ra}/{\la D_1\ra}=(13.8\pm 2.8)\%$ is in good agreement
with DELPHI result Eq.(\ref{apower})\footnote{SMC data
    \cite{bravardis99} yield an opposite sign,
    $\frac{\la H_1^\perp\ra}{\la D_1\ra}=-(10\pm 5)\%$,
    however, seem less reliable.}.
The errors are the statistical errors of the HERMES data.

The approach can be applied to predict azimuthal asymmetries in pion and kaon
production off {\sl a longitudinally polarized deuterium} target, which are
under current study at HERMES.
The additional assumption used is that ${\la H_1^{\perp K}\ra}/{\la D_1^K\ra}
\simeq{\la H_1^{\perp\pi}\ra}/{\la D_1^\pi\ra}$. The predictions are shown in
Fig.3. The "data points" estimate the expected error bars.
Asymmetries for $\bar K^0$ and $K^-$ are close to zero in our approach.
\begin{figure}[t!]
\begin{tabular}{ll}
\hspace{-0.5cm}\epsfig{figure=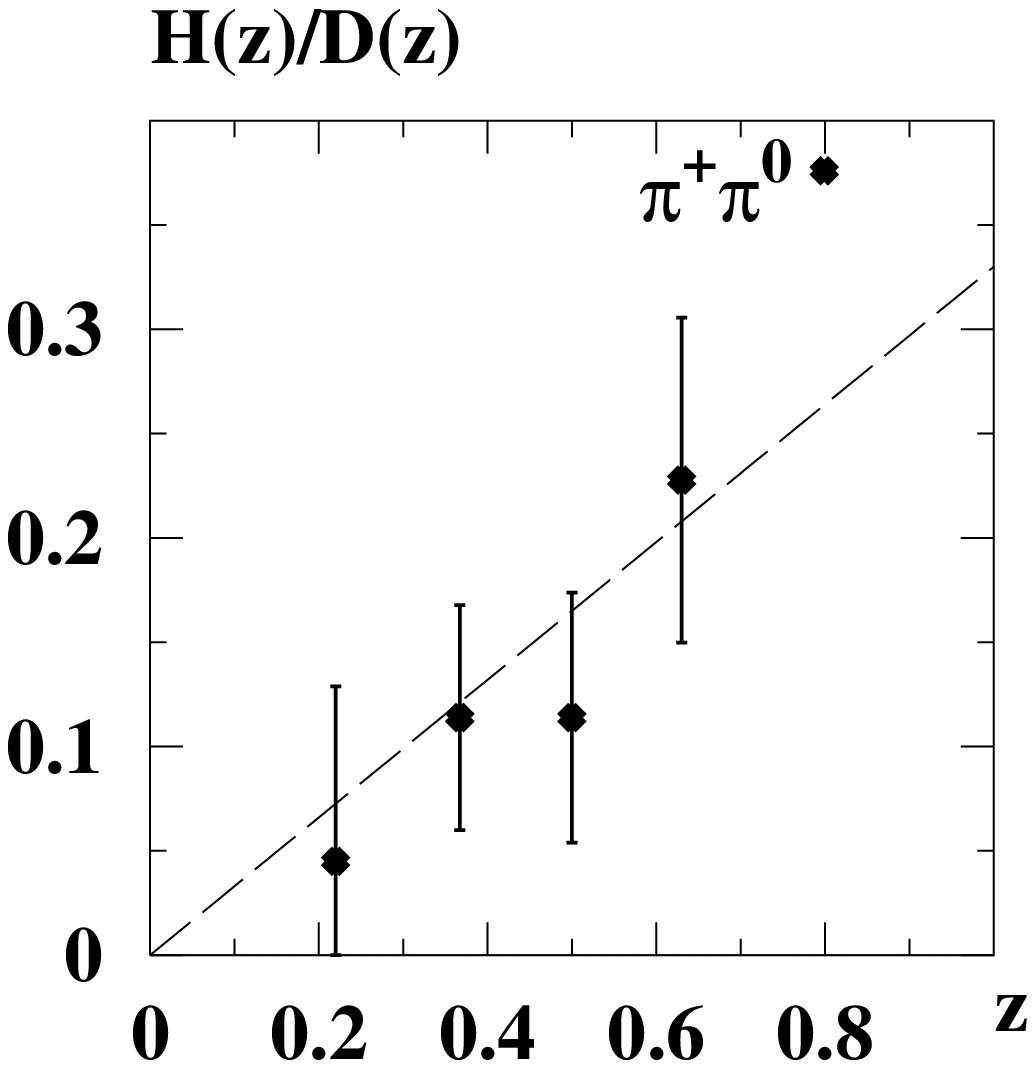,width=4cm,height=3.5cm}
    &
{\hspace{0.5cm}\epsfig{figure=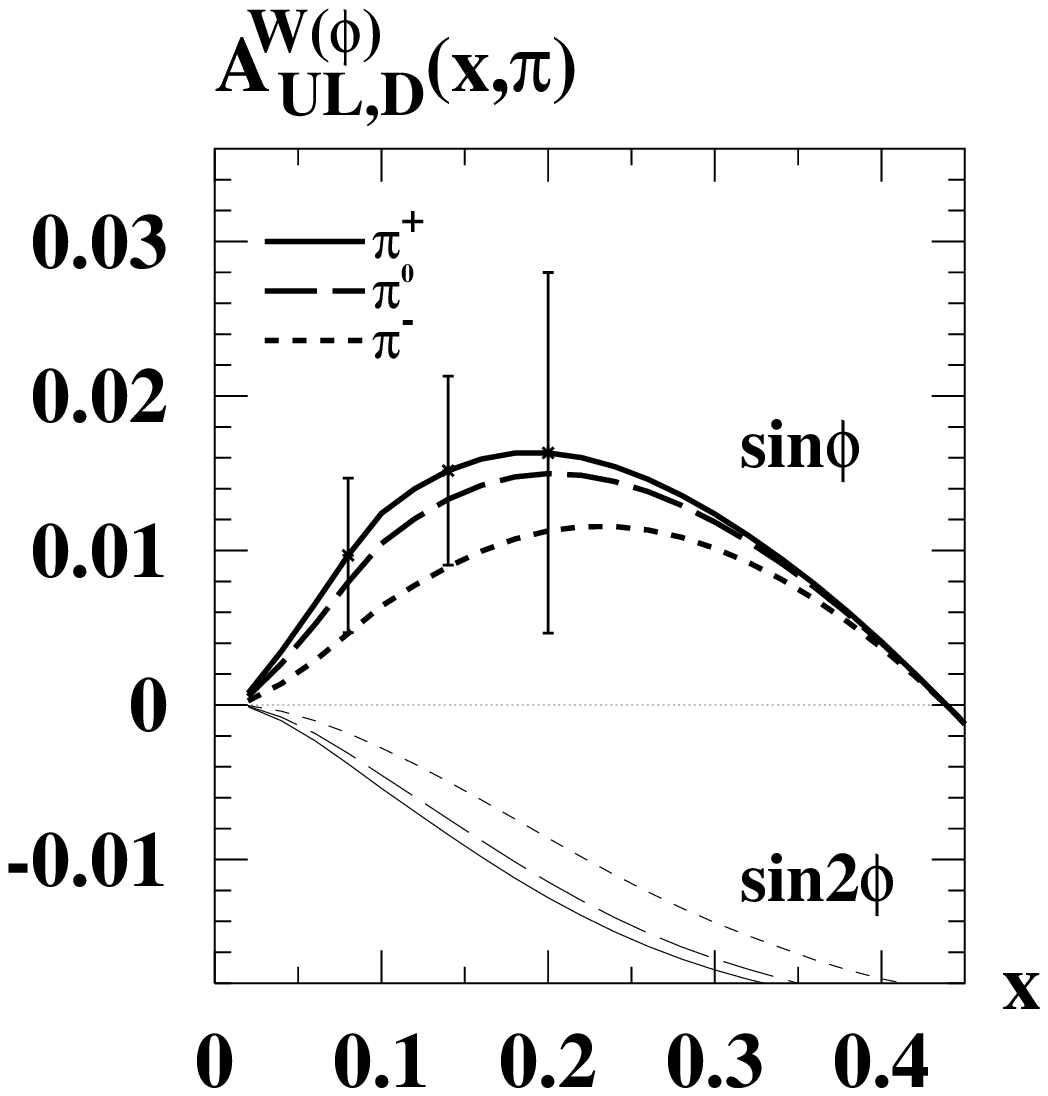,width=4cm,height=3.5cm}
\hspace{-0.7cm}\epsfig{figure=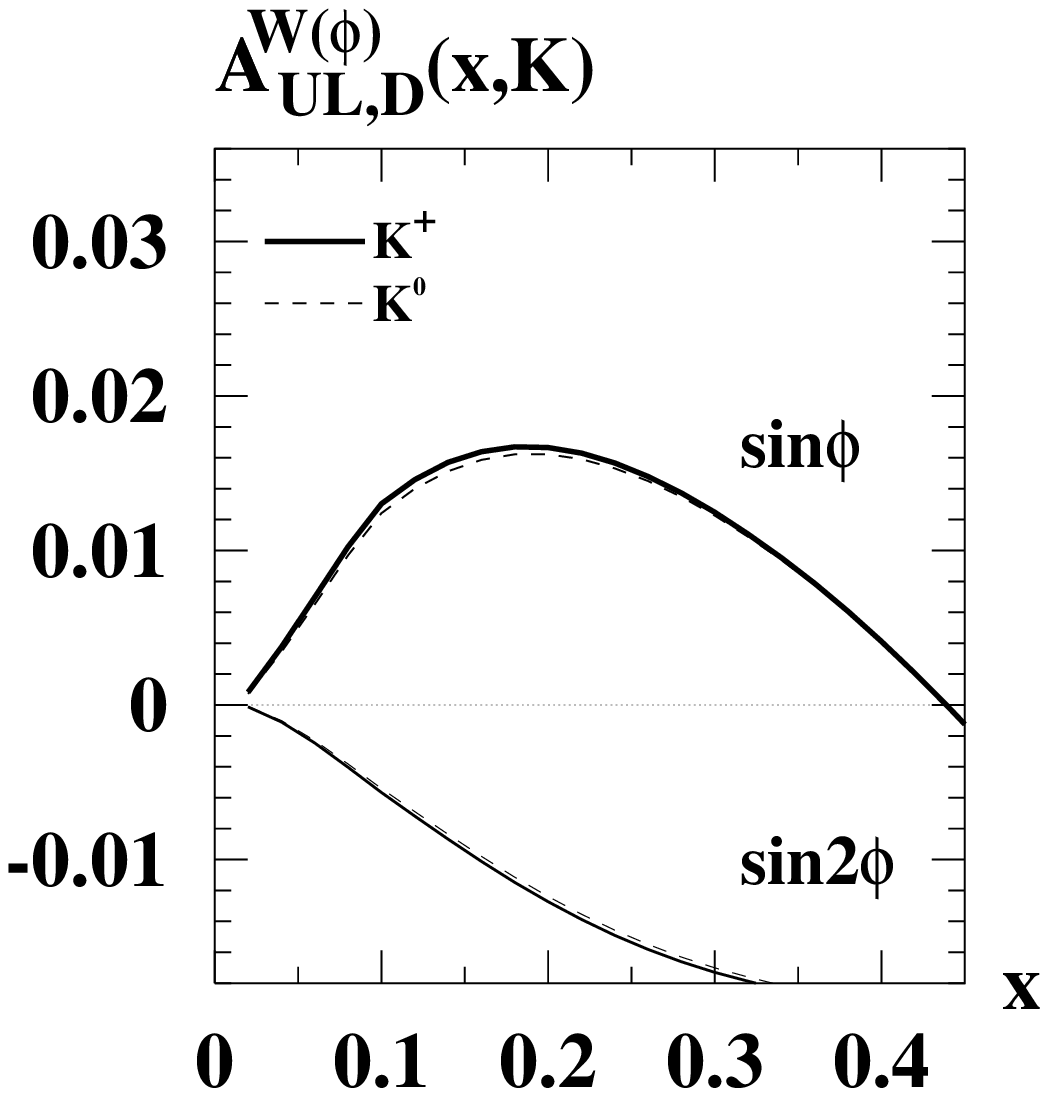,width=4cm,height=3.5cm}}
    \\
  \begin{minipage}{4.5cm}{\footnotesize Fig.2.
  $H_1^\perp/D_1$ vs. $z$, as extracted from HERMES data
  for $\pi^+$ and $\pi^0$ production \cite{hermes,hermes-pi0}.}\end{minipage}
    &
  \hspace{1cm}
  \begin{minipage}{7cm}{\footnotesize Fig.3.
  Predictions for $A_{UL}^{\sin\phi}$, $A_{UL}^{\sin2\phi}$ from a
  deuteron target for HERMES. Asymmetries for $\bar K^0$, $K^-$ are
  close to zero in our approach.}\end{minipage}
\end{tabular}
\end{figure}

Interestingly all $\sin\phi$ asymmetries change sign at $x\sim0.5$
(unfortunately the HERMES cut is $x<0.4$).
This is due to the negative sign in Eq.(\ref{AUT}) and the
harder behaviour of $h_1(x)$ with respect to $h_L(x)$.
This prediction however is sensitive to the favoured
fragmentation approximation.

We learn that transversity could be measured also
with a {\sl longitudinally} polarized target, e.g. at COMPASS,
simultaneously with $\Delta G$.

\section{Extraction of $e(x)$ from $A_{LU}^{\sin\phi}$ asymmetry at CLAS}

\begin{wrapfigure}{HR}{4.9cm}
  \vspace{-1.2cm}
  \mbox{\epsfig{figure=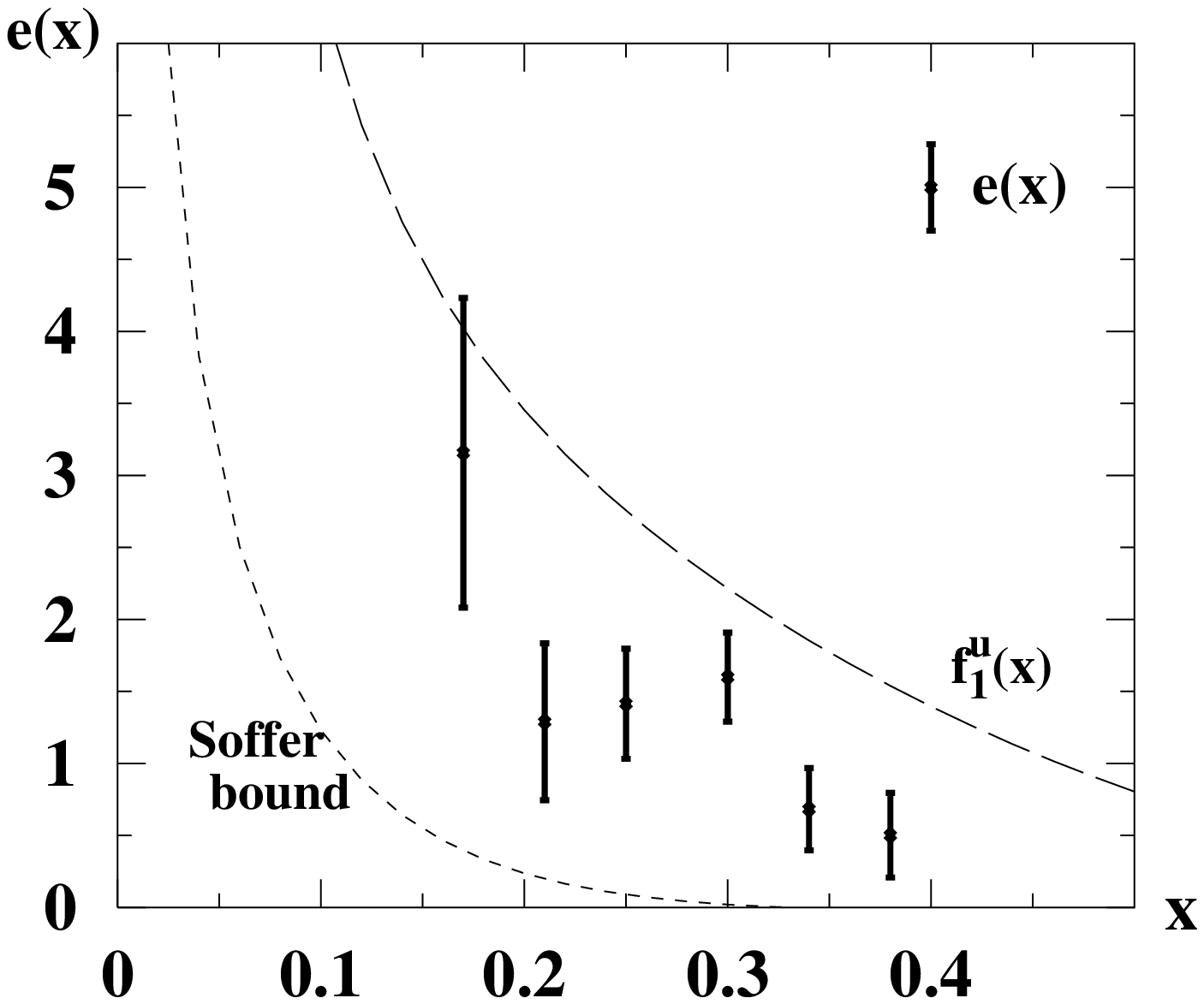,width=4.9cm,height=4.9cm}}
    {\footnotesize Fig.4. The flavour combination
    $e(x)\!=\!(e^u\!+\!\frac{1}{4}e^{\bar d})(x)$,
    with error bars due to statistical error of CLAS data,
    vs. $x$ at $\la Q^2\ra \!=\! 1.5\,{\rm GeV}^2$.
    For comparison $f_1^u(x)$ and the twist-3 Soffer bound are shown.}
  \vspace{-0.5cm}
  \end{wrapfigure}
Very recently the $\sin\phi$ asymmetry of $\pi^+$ produced by scattering of
polarized electrons off unpolarised protons was reported by CLAS collaboration
\cite{ALU}.
This asymmetry is interesting since it allows to access
the unknown twist-3 structure functions $e^a(x)$ which
are connected with nucleon $\sigma$-term:
\be\label{sigma}
    \int_0^1\di x\sum_ae^a(x)=\frac{2\sigma}{m_u+m_d}
    \approx 10 \; .\ee
The asymmetry is given by \cite{Mulders:1996dh}
\be
    A_{LU}^{\sin\phi}(x)\propto \frac{M}{Q} \;
    \frac{\sum_a e_a^2e^a(x)\la H^{\perp a/\pi}_1\ra}
    {\sum_a e_a^2 f^a_1(x)\la D_1^{a/\pi}\ra}  \;.\ee
Disregarding unfavored fragmentation and using the Collins analysing
power extracted from HERMES in Sect.\ref{AUL-HERMES}, which yields for
$z$-cuts of CLAS ${\la H^{\perp\pi}_1\ra}/{\la D_1^\pi\ra}=0.20 \pm 0.04$, we
can extract $e^u(x)+\frac14e^{\bar d}(x)$. The result is presented in Fig.4.
For comparison the Soffer lower bound \cite{Soffer:1995ww}
from twist-3 density matrix positivity,
$e^a(x)\ge 2|g_T^a(x)|-h_L^a(x)$,\footnote{For
    $g_T^a(x)$ we use the Wandzura-Wilczek approximation
    $g_T^a(x)=\int_x^1\di\xi\,g_1^a(\xi)/\xi$ and neglect consistently
    $\widetilde{g}_2^a(x)$ which is strongly suppressed in the
    instanton vacuum \cite{Balla:1997hf}.
    For $h_L^a(x)$ we use the analogous approximation, as described
    in Sect.3.}
and the unpolarized distribution function $f_1^u(x)$ are plotted.
One can guess that the large number in the sum rule Eq.(\ref{sigma})
might be due to, either a strong rise of $e(x)$ in the small $x$
region, or a $\delta$-function at $x=0$ \cite{Burkardt:2001iy}.

\vspace{0.2cm}
\noindent{\footnotesize
A.~E. is supported by RFBR grant 00-02-16696,
INTAS-00/587 and Heisenberg-Landau Programm, and
P.~S. by the contract HPRN-CT-2000-00130 of the European Commission.}

\end{document}